\documentclass[fleqn, usenatbib]{mnras}

\usepackage{amsmath,amssymb, graphicx}
\usepackage{epstopdf}

\usepackage[caption=false]{subfig}
\usepackage{enumitem}

\usepackage{natbib}
\usepackage{comment}
\usepackage{color}

\date{Accepted XXX. Received YYY; in original form ZZZ}

\pubyear{2020}

\newcommand{\jn}{\textcolor{black}}
\newcommand{\jnn}{\textcolor{black}}

\newcommand{\jnr}{\textcolor{black}}

\newcommand{\al}{\textcolor{black}}

\newcommand{\mpon}{\textcolor{black}}

\newcommand{\Lsi}{\lambda_{\mathrm{si}}}
\newcommand{\Lse}{\lambda_{\mathrm{se}}}
\newcommand{\Ex}{E_{\mathrm{x}}}
\newcommand{\Ey}{E_{\mathrm{y}}}
\newcommand{\Ez}{E_{\mathrm{z}}}

\newcommand{\Bz}{B_{\mathrm{z}}}

\newcommand{\Bo}{B_{\mathrm{0}}}

\newcommand{\Wci}{\Omega_{\mathrm{ci}}}
\newcommand{\Wce}{\Omega_{\mathrm{ce}}}

\newcommand{\degree}{^{o}}

\newcommand{\gsh}{\gamma_{\rm sh}}

\title[Mildly relativistic shocks in e$^{-}$p plasmas II]{Mildly relativistic magnetized shocks in electron--ion plasmas -- II. Particle acceleration and heating}

\author[Ligorini et al.]{Arianna Ligorini,$^{1}$
Jacek Niemiec,$^{1}$\thanks{E-mail: jacek.niemiec@ifj.edu.pl}
Oleh Kobzar,$^{2}$
Masanori Iwamoto,$^{3}$
Artem Bohdan,$^{4}$
\newauthor 
Martin Pohl,$^{4,5}$
Yosuke Matsumoto,$^{6}$
Takanobu Amano,$^{7}$
Shuichi Matsukiyo,$^{3}$
\newauthor
and Masahiro Hoshino$^{7}$
\\
$^{1}$Institute of Nuclear Physics Polish Academy of Sciences, PL-31342 Krakow, Poland\\
$^{2}$Astronomical Observatory of the Jagiellonian University, PL-30244 Krakow, Poland\\
$^{3}$Faculty of Engineering Sciences, Kyushu University, Kasuga, Fukuoka, 816-8580, Japan\\
$^{4}$DESY, 15738 Zeuthen, Germany\\
$^{5}$Institute of Physics and Astronomy, University of Potsdam, 14476 Potsdam, Germany\\
$^{6}$Department of Physics, Chiba University, 1-33 Yayoi, Inage-ku, Chiba 263-8522, Japan\\
$^{7}$Department of Earth and Planetary Science, University of Tokyo, 7-3-1 Hongo, Bunkyo-ku, Tokyo 113-0033, Japan\\
}

\begin{document}
\label{firstpage}
\pagerange{\pageref{firstpage}--\pageref{lastpage}}
\maketitle

\begin{abstract}
Particle acceleration and heating at mildly relativistic magnetized shocks in electron-ion plasma are investigated with unprecedentedly high-resolution two-dimensional particle-in-cell simulations that include {ion-scale shock rippling. Electrons are super-adiabatically heated at the shock, and most of the energy transfer from protons to electrons takes place at or downstream of the shock. We are the first to demonstrate that shock rippling is crucial for the energization of electrons at the shock. They remain well below equipartition with the protons. The downstream electron spectra are approximately thermal with a limited supra-thermal power-law component.} Our results are discussed in the context of wakefield acceleration and the modelling of electromagnetic radiation from blazar cores.
\end{abstract}


\section{Introduction}
High-energy charged particles produce intense nonthermal emission {that is observed from many astronomical objects in the Universe. They are also copiously registered at Earth as cosmic rays (CRs) with energy reaching $10^{21}$ eV, and possibly even higher at their sources}. The acceleration of these particles is one of the unsolved questions in astrophysics. 

Ultra-high-energy cosmic rays (UHECRs) with energies in excess of $\sim10^{18}$ eV are often {associated with} relativistic shocks formed in jets of active galactic nuclei (AGN) and/or gamma-ray bursts (GRBs). Blazar jets emit strong nonthermal synchrotron and inverse-Compton radiation {across the entire electromagnetic spectrum, which provides evidence for} electron acceleration to ultrarelativistic energies in these sources. High-energy electrons are most likely also responsible for the inverse-Compton component of the GRB jet's afterglow emission at sub-TeV $\gamma$-ray energies \citep{2019Natur.575..455M,2019Natur.575..459M,2019Natur.575..464A}. {Observations of high-energy neutrinos from the flaring blazar TXS 0506+056 indicate that relativistic jets also produce CR hadrons} \citep{2018Sci...361.1378I}. 

{Extragalactic jets probably harbour shocks whose Lorentz factor, $\gamma_{\rm sh}$, spans the range from near unity to a few hundred.}
Diffusive shock acceleration (DSA) may not work at ultra-relativistic ($\gamma_{\rm sh} \gg 1$) magnetized outflows, on account of inherent superluminal conditions at such shocks, in which particle diffusion across the magnetic field is suppressed \cite[e.g.][]{begelman1990,2006ApJ...650.1020N}. Particle-in-cell (PIC) simulation studies confirmed that ultra-relativistic shocks cannot be efficient particle accelerators through DSA-like processes unless the plasma magnetization is very small ($\sigma\lesssim 10^{-3}$, where $\sigma$ is the ratio of the Poynting flux to the kinetic energy flux) or the shock is subluminal \citep[][see also recent review by \cite{2020PrPNP.11103751P}]{2009ApJ...695L.189M,sironi2009,sironi2011}.  

As an alternative to the DSA model, {collective wave-particle interactions} can lead to efficient particle acceleration at superluminal magnetized relativistic shocks. Large-amplitude coherent X-mode waves are generated by the synchrotron maser instability (SMI) driven by the particles reflected off the shock-compressed magnetic field. {Upstream of the shock they form a} so-called precursor wave, {which has been demonstrated with particle-in-cell (PIC) simulations for} ultrarelativistic shocks ($\gsh \geq 10$) in one-dimensional (1D) \citep[e.g.][]{langdon1988,hoshino1991,gallant1992,hoshino1992,amato2006,plotnikov2019} and \emph{high-resolution} two-dimensional (2D) {studies} \citep[][\jnr{see also \citet{plotnikov2018}}]{iwamoto2017,iwamoto2018,iwamoto2019,2020MNRAS.499.2884B}. {Strong precursor waves in electron-ion plasma induce in their wake} large-amplitude longitudinal electrostatic oscillations \citep{lyubarsky2006,hoshino2008}. {Nonlinear collapse of the Langmuir waves generates nonthermal electrons and ions in a way similar} to wakefield acceleration \citep[WFA,][]{hoshino2008}, known from studies of laboratory plasma \citep[e.g.][]{1979PhRvL..43..267T,kuramitsu2008}. We recently demonstrated that {ultrarelativistic shocks with high electron magnetization in electron-ion plasma can accelerate electrons and ions in the turbulent wakefield to power-law spectra with index} close to $2$ \citep{iwamoto2019}. 

{Efficient heating of electrons to equipartition with the ions was reported by \citet{sironi2011} to happen before the flow reaches the shock. This} strong electron-ion coupling was argued to operate only for shocks with $\gsh\gtrsim (m_i/m_e)^{1/3}$, where $m_i/m_e$ is the ion-to-electron mass ratio \citep{lyubarsky2006}. PIC simulations of shocks with $\gsh \lesssim 10$ {that violate that condition showed very weak or no wakefield and little energy transfer from ions to electrons \citep{lyubarsky2006,sironi2011}.} However, these simulations were performed at low numerical resolution, whereas high resolution is required to properly account for the upstream wave generation \citep{iwamoto2017}.   

This is the second of two articles in which we study mildly-relativistic strictly perpendicular shocks in electron-ion plasma with unprecedentedly high-resolution and large-scale 2D PIC simulations, that account for ion-scale shock rippling. In \citet[][hereafter Paper I]{2020MNRAS.tmp.3691L}, 
we presented the electromagnetic structure of a shock with the Lorentz factor {$\gamma_{\rm sh}\simeq 3.3$} in plasma with total magnetization $\sigma=0.1$. {These parameters 
\jnn{may be relevant}
for internal shocks in AGN jets \citep[e.g.][]{ghisellini2005}. Here we discuss particle acceleration and heating at the shock to estimate the efficiency of WFA and electron-proton coupling in the mildly-relativistic regime.}
{Energy transfer between protons and electrons} at mildly-relativistic shocks is of importance in models of synchrotron and inverse-Compton emission from blazar jets. {Very strong coupling of electrons and ions would have consequences for the radiation spectra that impose quite strong constraints on the location of the emission sites in the jet and on the pair content \cite[e.g.][]{sikora2013}. Weak coupling for mildly relativistic conditions may relax some of those constraints.} 

\section{Simulation setup} 
\begin{figure}
\begin{center}
\includegraphics [width=0.99\linewidth] {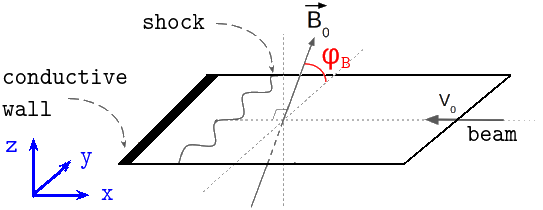}
\end{center}
\caption{Illustration of the simulation setup (adapted from Paper~I).}
\label{fig:setup}
\end{figure}

{Results presented in this paper are based on the same PIC simulations as in Paper I. Here we briefly reiterate the simulation setup that is illustrated in Fig.~\ref{fig:setup}. Initially cold, $v_{\rm th,beam}=0$, electron-ion plasma is reflected off a conducting wall at the left side of the simulation box. Its interaction with the incoming particles provides a shock moving in $+\boldsymbol{\hat{x}}$ direction.}
We use a modified version of the relativistic electromagnetic PIC code TRISTAN \citep{buneman1993} with MPI-based parallelization \citep{niemiec2008} and the option to trace individual particles.

We perform simulations in 2D3V in the $x-y$ plane, in which we track all three components of particle momenta and electromagnetic fields. 
We probe two configurations of the large-scale perpendicular magnetic field, $\boldsymbol{B_0}$, with respect to the simulation plane -- the \emph{out-of-plane} orientation with $\varphi_B=90\degree$ ($\boldsymbol{B_0}=B_{0z}\boldsymbol{\hat{z}}$) and the \emph{in-plane} setup with $\varphi_B=0\degree$ ($\boldsymbol{B_0}=B_{0y}\boldsymbol{\hat{y}}$). Together with $\boldsymbol{B_0}$, the beam carries the motional electric field $\boldsymbol{E_0} = - \boldsymbol{\mathit v_0 \times B_{0}}$, where $\boldsymbol{v_0}$ is the beam flow velocity.

\begin{figure*}
\centering
\includegraphics [scale= 0.37] {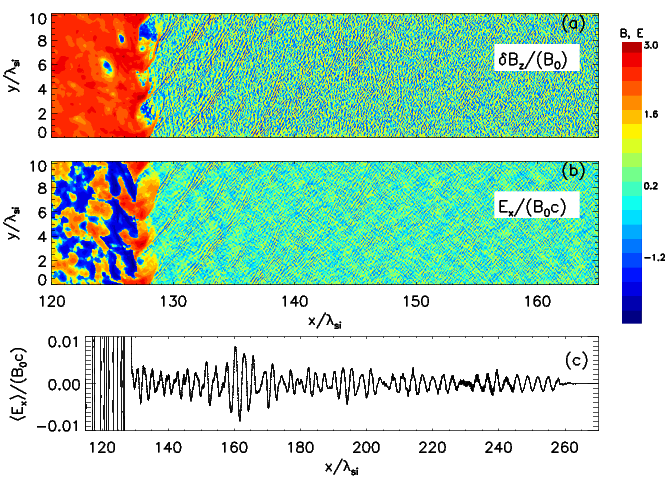}
\includegraphics [scale= 0.37] {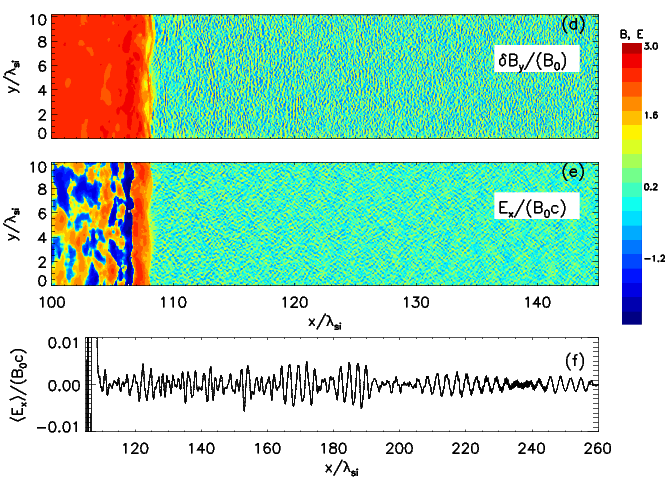}
\caption{\jn{Magnetic field oscillations of the} \al{X-mode waves and $E_x$ \jn{electric} field \jn{amplitude} at time $t\Wci = 84.3$. Logarithmic scaling is applied, which for the electromagnetic field is sign-preserving (e.g. $\mathrm{sgn}(B_{\text{X-mode}}) \cdot \{2+\log[\max(|B_{\text{X-mode}}|/B_{0},10^{-2})]\}$), so that field amplitudes below $10^{-2}B_{0}$ are not resolved. Panels (a) and (b) refer to the case with $\varphi_B=90\degree$, while panels (d) and (e) \jnr{refer} to $\varphi_B=0\degree$. Panels (c) and (e) show the respective transversely averaged profiles of the electric field, $\langle\Ex\rangle$, upstream of the shock.}}
\label{fig:shock_str}
\end{figure*}

For the beam Lorentz factor $\gamma_{0}=2.03$, the resulting Lorentz factor of the shock is $\gamma_{\rm sh}\simeq 3.3$ in the upstream rest frame. The total plasma magnetization is $\sigma = 0.1$, where $\sigma = B_{0}^2/(\mu_{0} N_{i}(m_{e} +m_{i})\gamma_{0}c^2)$, with the permeability of free space, $\mu_{0}$, and the upstream ion density, $N_{i}\, (=N_e)$, in the simulation (downstream) frame \citep{hoshino1992}. {The magnetization of electrons, $\sigma_{e}$, and ions, $\sigma_{i}$, are defined as} $\sigma_l = B_{0}^2/(\mu_{0} N_{l}m_{l}\gamma_{0}c^2), \, l=e,i$, so that $1/\sigma = 1/\sigma_{e} + 1/\sigma_{i}$. We assume a reduced ion-to-electron mass ratio of $m_{i}/m_{e}$ = 50, for which the electron magnetization is $\sigma_e\simeq 5.1$.

The final size of our simulation box, that expands in $x$-direction, is $L_x\times L_y=160000\times 5760\Delta^2$, where $\Delta$ is the grid cell size. We use the ion skin depth, $\Lsi = c/\omega_{\rm pi}$, as the unit of length, and $L_x\times L_y\simeq 283\times 10\,\Lsi^2$. Here, the ion plasma frequency is $\omega_{\rm pi}=\sqrt{e^2N_i/\gamma_0\epsilon_0m_{i}}$, where $e$ is the electron charge, and $\epsilon_0$ is the vacuum permittivity. The unit of time is the relativistic ion cyclotron frequency, $\Wci = (e\Bo)/(m_i \gamma_0)$. The simulation time is $t_{\rm max}\Wci=84.3$. We also ran a complementary 1D simulation up to $t_{\rm max}\Wci=163.1$. 

{The numerical resolution is set to $\Lse = 80\Delta$, where $\Lse=\sqrt{m_e/m_i}\Lsi$ is the electron skin depth.} This unprecedentedly high resolution is necessary to avoid artificial damping of the precursor waves.%
\footnote{\jnr{As discussed in \citet{iwamoto2017}, numerical damping of high-frequency precursor waves may result from the application of digital filtering, used to suppress the numerical short-wavelength Cherenkov modes. Precursor waves must therefore be resolved on scales larger than the wavelengths on which the filters operate. A~specific resolution depends on the numerical PIC model used and must be determined through test runs.}} 
The ion skin depth is thus $\Lsi\simeq 566\Delta$. We also set the number of particles per cell to $N_{\rm ppc}=10$ per species. {All parameter values have been selected after extensive tests} that are described in detail in Paper I.

\section{Electromagnetic shock structure \label{sec:structure}}
{We briefly summarize the shock structure that is in detail discussed in Paper~I.} As a point of reference, Fig.~\ref{fig:shock_str} shows normalized {maps of \al{the magnetic \jn{field oscillations associated with} the X-mode waves}, $\Ex$ fluctuations,} and profiles of transversely averaged electric field, $\langle\Ex\rangle$, at time $t=t_{\rm max}$ for the simulation runs with $\varphi_B=90\degree$ (left) and $\varphi_B=0\degree$ (right). For each magnetic-field configuration the SMI operates in agreement with theoretical predictions and produces coherent emission of upstream-propagating electromagnetic waves. {For out-of-plane magnetic field these precursor waves are entirely of the X-mode type, with fluctuating magnetic field along $\boldsymbol{B_0}$, whereas for 
in-plane magnetic field mode conversion provides also O-mode waves.} The strength of the waves, {$\delta B/B_0\simeq 0.19$ for $\varphi_B=90\degree$ and $\delta B/B_0\simeq0.15$ for $\varphi_B=0\degree$,} is much smaller than at high-$\gamma_{sh}$ shocks, where $\delta B/B_0\gg 1$, {but they persist and have comparable amplitudes as in a 1D simulation. Shock-front corrugation heavily influences the upstream plasma and the structure of downstream turbulence. In particular, it amplifies the precursor waves and counters suppression by }the inhomogeneous shock surface and the high temperature of the inflowing plasma. {The driver of shock rippling depends on the magnetic-field orientation} -- it is the process described by \citet{burgess2007} for $\varphi_B=90\degree$ and the \mpon{Alfv\'en Ion Cyclotron (AIC)} temperature-anisotropy instability for $\varphi_B=0\degree$. The wavelength of the ripples is $\lambda_{\rm rippl}\simeq 3.3\Lsi$ and $\lambda_{\rm rippl}\simeq 5\Lsi$, respectively.  
{For out-of-plane magnetic field the precursor waves are on average emitted obliquely to the shock normal, which reflects wave emission in a direction normal to the local tangent to the ripplings' arcs and the effects of retardation and aberration, as the ripples move with $v_{\rm rippl}\approx 0.8c$ along the shock surface.}
The AIC-generated shock front corrugations in the in-plane case are of slightly lower amplitude, and the waves are mostly propagating along the shock normal with only a weak oblique component.

In all simulations we observe density filaments upstream of the shock that are generated by the parametric filamentation instability \citep{kaw1973,drake1974}. Likewise, in each case electrostatic wakefield of moderate amplitude is excited in the upstream plasma. {Using} the so-called strength parameter, $a=e\delta E/m_ec\omega$, where $\delta E$ is the  electric-field amplitude and $\omega$ is the wave frequency \citep{kuramitsu2008}, {we find with significant fluctuations} $a\simeq 0.15-0.30$ for $\varphi_B=90\degree$ and $a\simeq 0.2$ for $\varphi_B=0\degree$.

\section{Particle Heating and Acceleration}
\label{sec:spectra}

\subsection{Out-of-plane magnetic field: \jn{$\varphi_B=90\degree$}}

\subsubsection{Electron interactions with upstream waves \label{ele_upstream}}

\begin{figure}
\centering
\includegraphics [scale= 0.4] {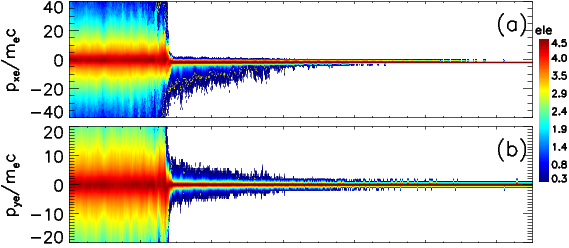}
\includegraphics [scale= 0.4] {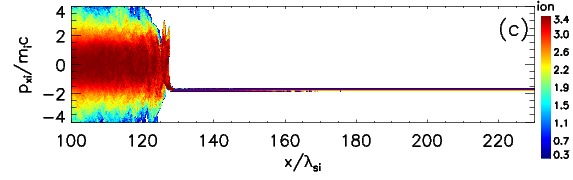}
\includegraphics [scale= 0.4] {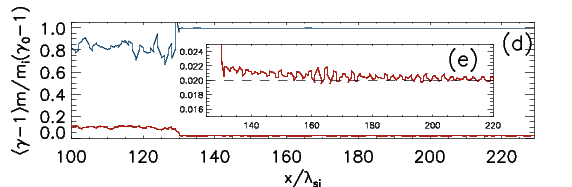}
\caption{\jn{Phase-space distribution in $\log(p/mc)$ of electrons and ions across the shock at time $t\Wci = 84.3$, averaged over the spatial coordinate $y$. From top to bottom, shown are the $x$- and $y$-components of electron momentum and the $x$-component of ion momentum.} {Panel (d) displays in blue for ions and in red for electrons the mean kinetic energy in units of that of far-upstream ions. Inserted is a rescaled kinetic-energy profile of electrons (e).}}
\label{fig:phase_sp}
\end{figure}

Fig.~\ref{fig:phase_sp} shows electron and ion phase-space distributions across the shock, as well as the mean particle kinetic energy, $\langle\gamma-1\rangle m_l c^2$. {Electrons streaming toward the shock are accelerated toward it and gradually heated, reaching $\gamma\approx {20}$ close to the shock. Panel e) indicates that some $50\Lsi$ ahead of the shock the mean electron energy commences a steady but slow growth that results from interactions with the wakefield. We know from Fig.~\ref{fig:shock_str} that further away the amplitude of the wakefield does not exceed $\langle\Ex\rangle/B_0c\approx 4\times 10^{-3}$, and closer than $50\Lsi$ to the shock it is only marginally larger than that.
Incoming electrons interacting with a Langmuir wave of such amplitude should show an acceleration-deceleration pattern in the mean electron energy that we can indeed see in Fig.~\ref{fig:phase_sp}(e).
There is little or no net gain in energy, but the coherent oscillations in velocity can be effectively regarded as heating to} the maximum energy \citep{hoshino2008}:
\begin{equation}
    \frac{\epsilon_{\rm max}}{\gamma_0 m_ec^2}\approx e \langle\Ex\rangle L \simeq \frac{\xi a^2}{\sqrt{1+\xi a^2}}(1+\beta_0)\simeq 0.1,
\end{equation}
{where $L\approx 1/k_L$ is the scale-length of the wakefield and $\xi=1/2$ is appropriate for a linearly polarized precursor wave. We inserted the peak} \jn{value of the strength parameter, $a=0.3$. This energy level}
is {compatible with the typical amplitude of electron energy fluctuations. The electrons with $\gamma\approx 20$ that we noted in Fig.~\ref{fig:phase_sp} are very few, do not represent the bulk, and are likely involved in nonlinear interactions with the wakefield.}

{In electron-ion plasma the wakefield is generated by the ponderomotive force on electrons exerted by the precursor wave that leads to collective motion and Langmuir waves \citep{hoshino2008}.}
\jn{Wakefield formation can be also understood as result of the parametric decay instability \citep[PDI; e.g.][]{kruer1988}.}
{Parametric instabilities are common wave-wave interactions that arise from wave coupling at a nonlinearity such as a pressure gradient. Frequency and wavenumber matching are consequences of energy and momentum conservation. In our case, \jn{nonlinear} stimulated Raman scattering turns a large-amplitude electromagnetic (pump) wave into a Langmuir wave and a scattered electromagnetic (light) wave. If the pump wave frequency is much larger than the plasma frequency, we have Forward Raman Scattering \jn{(FRS)}, and the scattered electromagnetic wave and the Langmuir wave propagate in the same direction as the pump wave. In \jn{subsequent stages} the scattered electromagnetic waves can further decay into other electromagnetic waves and Langmuir waves, and eventually broadband wave spectra can be generated. In the upstream plasma frame, the waves propagate outward, but in the simulation frame part of the electromagnetic and Langmuir waves can propagate in downstream direction \citep{hoshino2008}.
Fig.~\ref{fig:shock_str}(e) indicates a much higher wakefield amplitude close to the shock than in the far-upstream region, which reflects amplification of precursor-wave emission through shock rippling. 
We demonstrated in Section 3.2.5 of Paper I, that in the region extending as far as $(x-x_{\rm sh})/\Lsi\approx 70$ from the shock the wakefield moves toward the shock, which we interpret as a signature of nonlinear FRS, during which the wakefield can collapse and transfer energy to particle heat and bulk acceleration.}

{As at ultrarelativistic shocks, the downstream-propagating Langmuir waves can resonantly interact with the electrons through the so-called phase-slippage effect, boost them toward the shock, and create an asymmetric wing in the $p_x$ phase-space distribution. The Langmuir waves also facilitate shock-surfing acceleration (SSA) that would accelerate trapped electrons in the $y$-direction and cause the anisotropy in the $p_{\rm ye}$ distribution toward positive momenta that is evident in Fig.~\ref{fig:phase_sp}(b). The trajectory analysis in Section~\ref{sec:tracing} demonstrates energy gains and losses correlated with motion in $y$-direction.
Note that also the ions show similar disturbances in their phase-space distribution.}

{Despite the presence of coherent precursor waves, the bulk energy of electrons increases by only 5\% before they hit the shock (see Fig.~\ref{fig:phase_sp}e). Most of the energy transfer from ions to electrons takes place at the shock and in the near-downstream region, as we discuss next.}

\begin{figure}
\begin{center}
\includegraphics [width=0.99\linewidth] {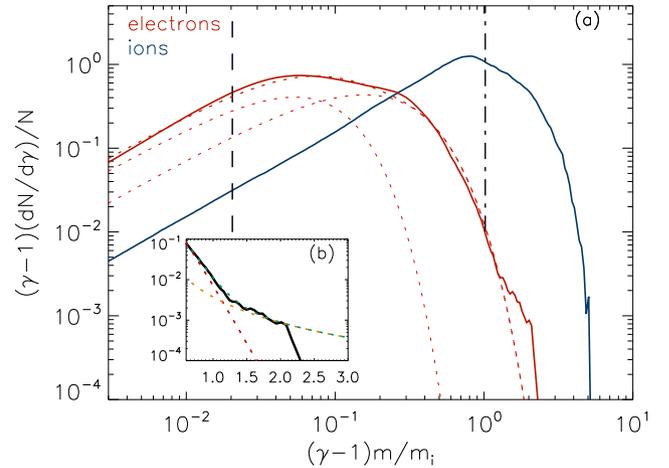}
\end{center}
\caption{{Energy spectra for electrons (red) and ions (blue) downstream of the shock ($x/\Lsi = 113-123$) at time $t\Wci = 84.3$. The energy axis is scaled with the particle mass, $m=m_e, m_i$. Vertical dashed and dash-dotted lines mark the initial kinetic energies of the electrons and ions, respectively. Also shown with thin red dotted lines are two relativistic 2D Maxwellians that together (thick dotted line) is an eyeball fit to the electron spectrum. Inset (b) highlights the high-energy part of the electron spectrum, in which the double Maxwellian (red), a power-law $\propto (\gamma-1)^{-2.1}$ (yellow), and their sum (green) are shown with dotted lines.}}
\label{fig:spectra}
\end{figure}

\begin{figure*}
\begin{center}
\includegraphics [width=0.99\linewidth] {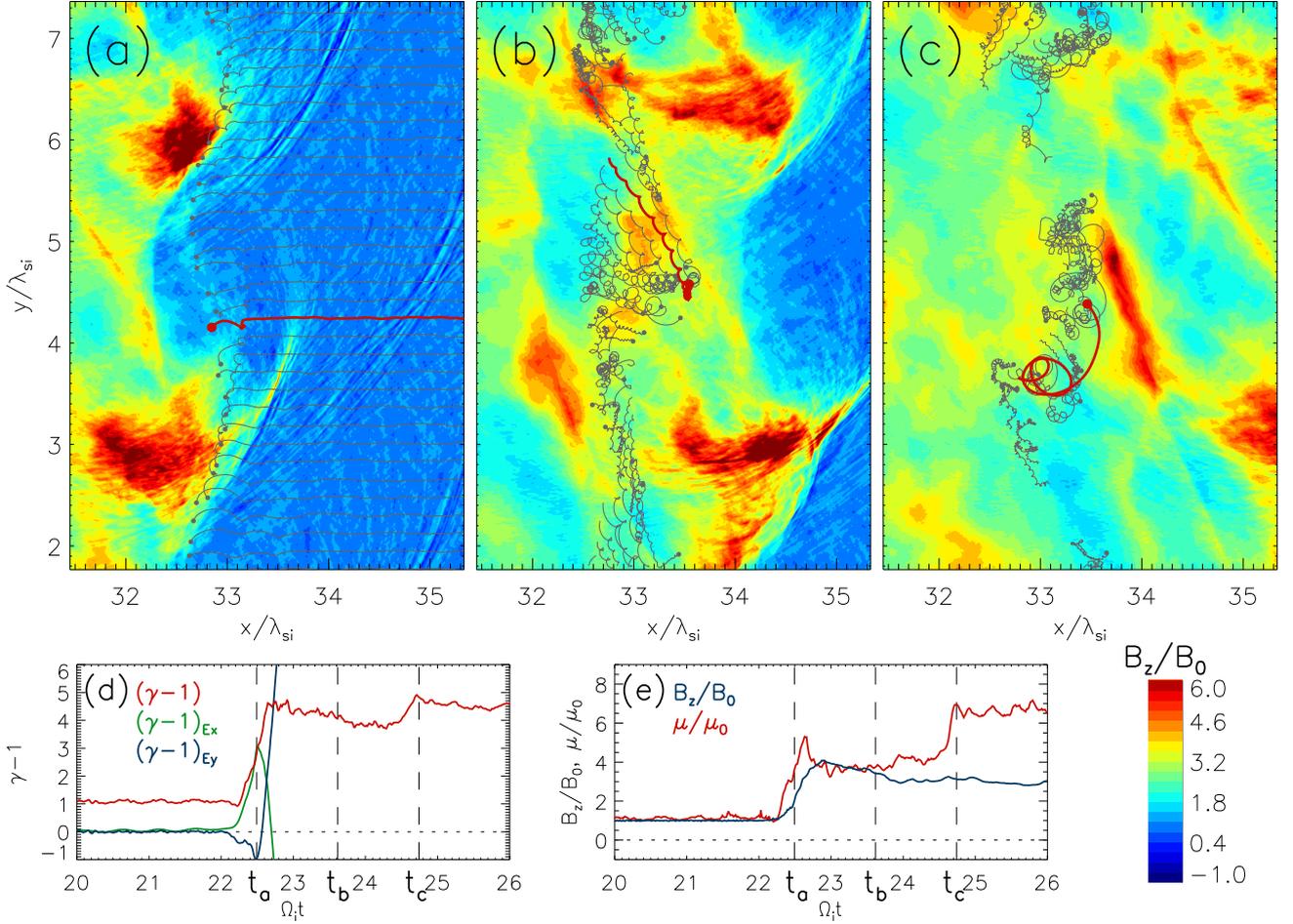}
\end{center}
\caption{{Electron trajectories ending at the circles on a map of normalized magnetic-field strength, $|B_z|/B_0$. The map is linearly scaled and refers to the end times $t_a$, $t_b$, and $t_c$, that are marked in panels (d) and (e). The trajectories span $0.3\Wci^{-1}$. In each panel a selected electron trajectory is plotted in red. Panel (d) shows the evolution of the average kinetic energy (red line) as well as the the $x$- (green line) and $y$- (blue line) components of the \emph{average} electric work (cf. Eq.~\ref{eq:work}). Panel (e) shows the averaged magnetic moment (red line) and the magnetic field, $B_z$ (blue line), along the trajectories.}}
\label{fig:tracing}
\end{figure*}

\subsubsection{Downstream particle spectra \label{sec:downstream_spectra}}
{Fig.~\ref{fig:spectra} shows kinetic-energy spectra of electrons and ions at time $t\Wci = 84.3$, taken in the near-downstream region at $x/\Lsi = 113 - 123$. The particle spectrum represents a steady-state distribution that does not appreciably change with time and distance from the shock.
Although some electrons reach energies exceeding the kinetic energy of upstream ions, \jn{on} average they have only 13\% of that of the ions. The two species do not reach the equipartition that is typically observed at ultrarelativistic shocks \citep[e.g.][]{lyubarsky2006,sironi2011}. 
Downstream of the shock ions and electrons carry $81.5\%$ and $10.6\%$ of the initial ion kinetic energy, respectively. In the simulation frame the mean kinetic energy in the upstream region should be similar to that downstream, and so about $8\%$ of the energy are available for transfer to electromagnetic waves and turbulence.}

{The ion spectrum tends towards a 2D relativistic Maxwellian distribution,
\begin{equation}
\frac{dN}{d\gamma}\propto \gamma \exp\left(-\frac{mc^2}{kT}\,\gamma\right).
\end{equation}
The high-energy particles with Lorentz factors up to $\gamma_i\approx 6$ bounced off the shock and typically perform a single loop in the upstream motional electric field, experiencing one \jn{shock-drift acceleration} (SDA) cycle.
The electron distribution is more asymmetric and best fitted with two Maxwellians. The hotter Maxwellian peaks at $(\gamma_e-1)m_e/m_i\simeq 0.2$, and the other one is about 6 times colder. The electron spectrum also shows a weak supra-thermal component at $(\gamma-1)/m_i\gtrsim 1$, that is compatible with a power-law of index $p\approx 2$, shown with the yellow dotted line in Fig.~\ref{fig:spectra}(b).}

\subsubsection{Energization processes \label{sec:tracing}}
{Although coherent precursor waves and wakefield exist upstream, they cause only little energy transfer from ions to electrons.
Nevertheless, superadiabatic particle heating is observed at the shock, and some supra-thermal electrons are produced.
On account of weak electron-ion coupling in the upstream region,
the ions enter the shock with much larger energy than do electrons, reach deeper in the shock-compressed magnetic field, and \mpon{cause charge separation producing} electrostatic field in upstream direction. With out-of-plane field, $\boldsymbol{B_0}=B_{0z}$, particle motion is constrained in the $x-y$ plane, and so the cross-shock electric potential cannot provide electron heating parallel to $B_0$.
In this section we analyze particle trajectories to study electron energization at the shock.}

Fig.~\ref{fig:tracing} illustrates the main stages of electron energization {as they cross the shock. The particles are selected in the far-upstream region at the same $x$-location, and they approach the shock ramp at the same time. 
Some trajectories are shown with thin grey lines in panels (a)-(c), and a single electron in each panel
is highlighted in red. The temporal evolution of the kinetic energy averaged over \emph{all} \jn{traced} particles, not only those visible in panels (a) to (c), is shown in red in panel (d), where we also plot the normalized work done by the $\Ex$ (green line) and $\Ey$ (blue line) electric-field components, 
\begin{equation}
(\gamma-1)_{E_i} = \frac{-e }{m_e c^2}\Big\langle\int_{t_0}^{t} dt'\  v_i(t')\, E_i(\mathbf{x}(t')) \Big\rangle\ ,
\label{eq:work}
\end{equation}
where $i=x,y$.
The local electric field is measured at the particle position and $t_0\Wci=20$. Finally, in panel (e) we plot in red the average of the magnetic moment, $\mu$, and the $B_z$ profile (blue line) along the particle trajectories. Here, $\mu=p_{\perp}/(2m_e|\boldsymbol{B_p}|)$, where $p_\perp$ is the transverse momentum of the particles in the local magnetic field, $\boldsymbol{B_p}$, and $\mu_0=\mu(t_0)$.}

\begin{figure}
\begin{center}
\includegraphics [width=0.99\linewidth] {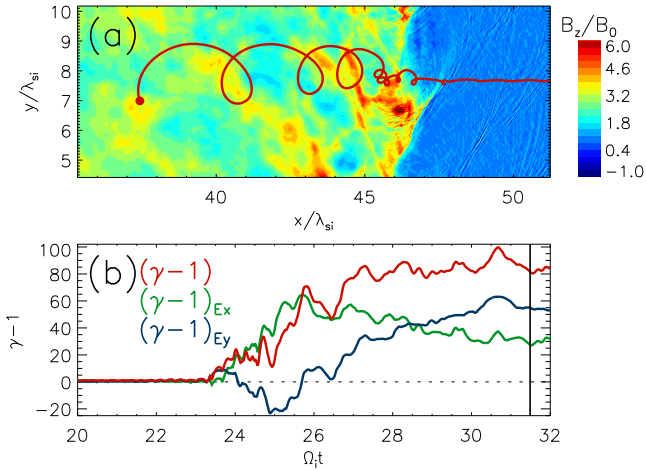}
\end{center}
\caption{Trajectory of a highly energetic electron overlaid {on a linear map of $\Bz$ magnetic field (a). The map corresponds to time $t\Wci = 23.6$, when the particle first interacts with the shock front. Panel (b) shows the evolution of the total kinetic energy and the electric work (cf. Eq.~\ref{eq:work}) as in Fig.~\ref{fig:tracing}, but here for a single electron. 
The black vertical line at time $t\simeq 31.5$  marks the end point of the trajectory.}}
\label{fig:tracing_max}
\end{figure}

{As electrons approaching the shock interact with the wakefield, far upstream most of the particles only oscillate in the electrostatic field but do not significantly gain energy. Closer to the shock the wakefield is stronger and the electrons may become decoupled from the bulk flow \jnn{\citep[compare][]{sironi2011}}. SSA kicks in, and the decoupled electrons are accelerated in the $y$-direction, which should be visible as anisotropy in the $p_{\rm ye}-x$ phase-space distribution. To be noted from Fig.~\ref{fig:tracing}(a) is that only a small fraction of particles are decoupled from the bulk flow and show small-amplitude oscillations in their trajectories. }

{A significant impact of the upstream waves can be observed just in front of the shock.
Fig.~\ref{fig:tracing}(a) shows strong waves emitted by a shock ripple that form an arc-like feature in the $\Bz$ distribution and corresponding $\Ex$ and $\Ey$ components (not shown). The $\Ex$ wave field is strong enough to effectively stop an electron and decouple it from the bulk flow. This causes wiggles in the trajectory, well visible for the highlighted electron at $(x/\Lsi,y/\Lsi)\approx(33.2,4.2)$ and the particles below it at $y/\Lsi\approx 3.2-4$ that were smashed by the waves emitted by the lower ripple, and particles at $y/\Lsi\approx 5.9-7.4$, affected by the upper ripple. Note, that at time $t_a$ the affected electrons have already passed through the waves. After decoupling, the electrons gain significant energy by the $\Ex$ field and to a lesser degree by the motional $\Ey$ field. The strong electrostatic $\Ex$ field is a combination of the strong wakefield close to the ripple and the standard cross-shock field. 
The jump in the average magnetic moment (panel (e)) indicates that energization at this stage is nonadiabatic.}

This scenario of electron acceleration close to the shock is a crucial and necessary step for their subsequent energization. It fully relies on strong rippling of our mildly relativistic shock \mpon{and is, to our knowledge, 
described here for the first time.}

{Deeper in the shock, electrons are adiabatically heated while they undergo $\boldsymbol{E}\times\boldsymbol{B}$ drift in the $-\Ex\Bz\boldsymbol{\hat{y}}$ direction with the same speed as the shock ripples. The combination of
penetration and drifting of ions causes the charge-separation electric field in the ripples whose components have amplitudes $\Ex\approx\Ey\gg E_0$. In the simulation frame the electrons are dragged together with the ripples and are accelerated by the $\Ey$ field and decelerated by the $\Ex$ field. Hence the components of work $(\gamma -1)_{\Ex}$ and $(\gamma -1)_{\Ey}$ diverge at $t\gtrsim t_a$. The drifting of electrons is evident in Fig.~\ref{fig:tracing}(b).}

{Obviously, the behavior of individual particles may differ from that of the average particle. Electrons that decoupled from the bulk flow in the upstream region and started gyrating there may hit the shock with unfavorably low energy. After adiabatic heating at the shock \jn{they} will be advected downstream and form the thermal pool together with electrons that passed through the shock without interacting with the ripple structures.}

{Having non-adiabatically gained energy around time $t_a$, the average electron resides at the overshoot, and its further energy evolution is predominantly adiabatic. Fig.~\ref{fig:tracing} illustrates that around time $t_b$ the magnetic moment is conserved while the magnetic field becomes weaker, which results in a gradual energy loss due to decompression. Later, around $t\approx t_c$, the mean electron energy again increases, associated with a second jump of the magnetic moment. The particle trajectories in panel (c) show that the gyroradii of some electrons are comparable to the scale of the turbulent field, allowing resonant scattering off these waves. Essentially all electrons with Lorentz factor $\gamma_e\gtrsim 10$ can gain energy in this way and populate the high-energy wing of the spectrum downstream.}

{At times $t>t_c$ the average electron energy saturates, probably for want of large-scale turbulence with which they could interact. The turbulence exists only in a narrow zone downstream of the overshoot, and its scale is commensurate with the ripple wavelength.}

\begin{figure}
\centering
\includegraphics [scale= 0.4] {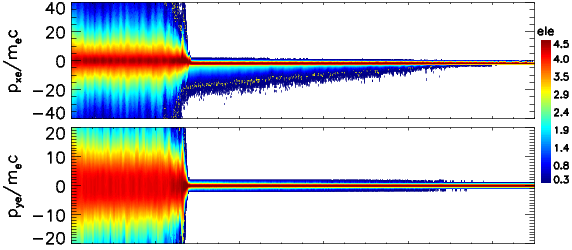}
\includegraphics [scale= 0.4] {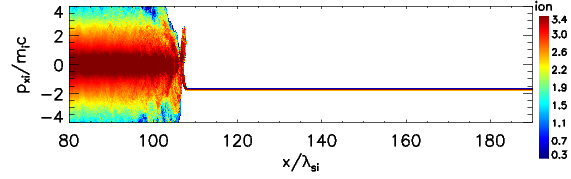}
\includegraphics [scale= 0.4] {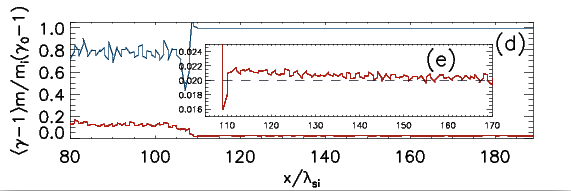}
\caption{\jn{Phase-space distributions of electrons and ions at time $t\Wci = 84.3$ {in the simulation with in-plane mean magnetic field. Also shown in panels (d) and (e) are profiles of the mean kinetic energy, red for electrons and blue for ions.}}}
\label{fig:phase_ip}
\end{figure}

The features discussed here for the average electron {can explain the downstream electron spectrum as shaped by a single population of particles that was energized in the shock.
The particles, that decoupled from the bulk flow in the upstream region and were energized through ripple-mediated processes, can experience further acceleration in resonant interactions with turbulence. A consequence is the smooth transition between the low-energy and high-energy component in the electron spectrum. Some energization at the shock can be described as a stochastic second-order Fermi process. An example is given by the particle shown in Fig.~\ref{fig:tracing_max}. After decoupling from the flow this electron experienced multiple instances of inelastic scattering in the shock transition, gaining energy on average and finally reaching $\gamma_e\approx 80$. Such particles form the supra-thermal power-law portion of the electron spectrum. 
Most of the scattering is less successful, and the electrons are in the bulk of the population. 
In our mildly relativistic magnetized shock, the layer of strong electromagnetic turbulence is relatively narrow, and stochastic scattering does not provide significant energy gain to the average electron.}

\jnn{Double-Maxwellian downstream electron distributions were reported before for ultrarelativistic superluminal shocks \citep{sironi2011}. However, their high-energy spectral component results from energization through wakefield in the shock upstream, that is not efficient in our mildly relativistic shock.
Here, the formation of the hotter Maxwellian and its supra-thermal component entirely results from electron interactions with turbulence at the shock and downstream, that is \mpon{related to} the shock ripples.}

\subsection{In-plane magnetic field configuration: $\varphi_B=0\degree$}
\label{sec:in-plane}

\subsubsection{Upstream particle-wave interactions and downstream spectra}
\label{sec:spectra_ip}

{For the in-plane case the phase-space distributions of electrons and ions and the mean kinetic-energy profiles presented in Fig.~\ref{fig:phase_ip} are very similar to those for the simulation with $\varphi_{\rm B}=90\degree$. The amplitudes of the precursor waves and the wakefield are compatible in both runs, and so essentially the same physical mechanisms of wave-particle interaction operate. In particular, electrons can be accelerated by strong wakefield propagating toward the shock, reach energies in excess of $\gamma_e\sim 20$, and develop anisotropy in the $p_{\rm xe}$ distribution. SSA works as well, although the corresponding $p_{\rm ze}$ anisotropy along the motional electric field cannot be observed in our 2D simulation. The bulk energy gain of electrons before they reach the shock is again about 5\% of their initial kinetic energy, and the electrons and ions are far from energy equipartition.}  

\begin{figure}
\begin{center}
\includegraphics [width=0.99\linewidth] {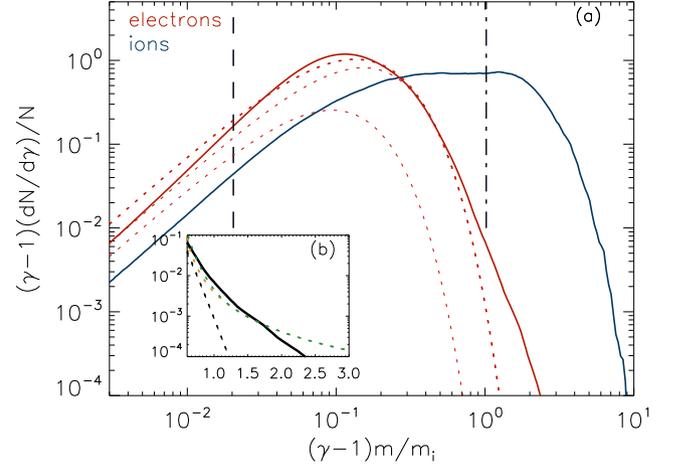}
\end{center}
\caption{{Spectra of electrons in red and ions in blue downstream of the shock ($x/\Lsi = 87-97$ at $x_{\rm sh}/\Lsi\approx 107$)
at time $t\Wci = 84.3$. The dotted line shows fits of relativistic 3D Maxwellians. Inset (b) is a close-up of the high-energy part of the electron spectrum.}}
\label{fig:spectra_ip}
\end{figure}

\begin{figure}
\begin{center}
\includegraphics [width=0.99\linewidth] {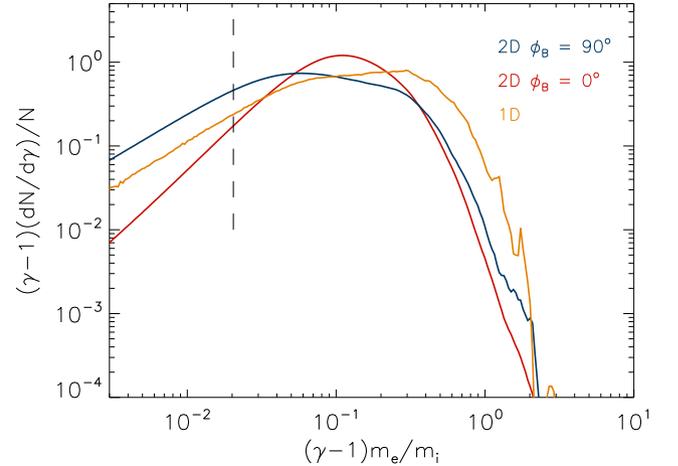}
\end{center}
\caption{Comparison of downstream electron spectra in 2D (blue line for $\varphi_{\rm B}=90\degree$, red line for $\varphi_{\rm B}=0\degree$) and 1D simulations (orange line). The spectra are calculated in a region $5\Lsi-15\Lsi$ downstream of the shock front.}
\label{fig:coupling}
\end{figure}

Fig. \ref{fig:spectra_ip} shows downstream particle spectra. \jn{As with out-of-plane magnetic field, the ions are in the process of thermalization, and the reflected particles undergo SDA at the shock.} 
The electron distribution is close to \jn{a combination of two 3D Maxwellians} {with moderately different temperature, each described by 
\begin{equation}
 \frac{dN}{d\gamma}\propto \gamma\,\sqrt{\gamma^2-1}\,\exp\left(-\frac{m_ec^2}{kT_e}\,\gamma\right).
\end{equation}}
\jn{A single 3D Maxwellian provides a slightly worse fit.} 
{Inset~(b) of Fig.~\ref{fig:spectra_ip} features a weak {supra-}thermal component with spectral index $p\approx 2$,} similar to that in
the out-of-plane simulation. 

\begin{figure*}
\begin{center}
\includegraphics [width=0.99\linewidth] {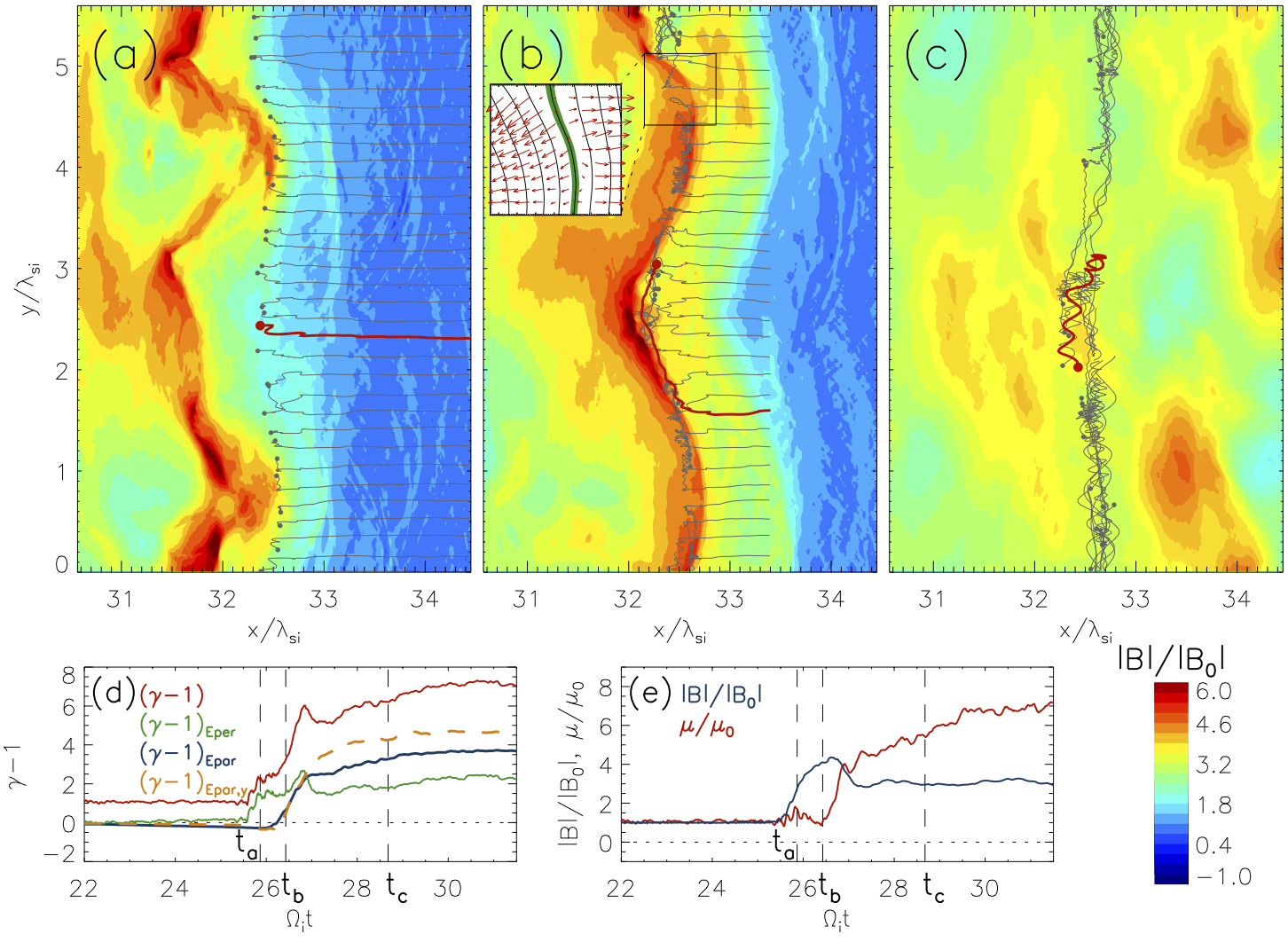}
\end{center}
\caption{Trajectories of electrons (a-c), the evolution of the average kinetic energy and electric work (d), and the magnetic moment and the magnetic-field profile (e) in the simulation with $\varphi_{\rm B}=0\degree$. The format is the same as in Fig.~\ref{fig:tracing}, except that the normalized magnitude of the magnetic field, $|B|/B_0$, is shown in panels (a-c) and (e), and the acceleration rates are split into components parallel (blue line) and perpendicular (green line) to the local magnetic field.   
For $(\gamma-1)_{E_{\rm par}}$ we also show its $y$-component in orange. The inset in panel (b) shows zoom-in of the fields configuration in a region marked with a square box, where red arrows present the in-plane electric field, black lines the contours of the $A_z$-component of the vector potential (displaying the in-plane magnetic field lines), and the green thick line shows the region with strong electric field parallel to the magnetic field.}
\label{fig:tracingIP}
\end{figure*}

{Downstream of the shock electrons and ions are far from energy equipartition, as in the setup with $\varphi_{\rm B}=90\degree$.} To facilitate a direct comparison, in Fig.~\ref{fig:coupling} we once again show the electron spectra for both 2D runs and also for a 1D simulation. 
{With in-plane field, the downstream electrons carry $ 12.6\%$ of the initial ion kinetic energy, the ions account for $77.2\%$, and the rest is available for fields and turbulence. The electron/ion energy ratio} is thus slightly larger in the in-plane case, but comparable in both 2D runs. In 1D simulations the coupling is slightly stronger, the electrons gain $16,8\%$ of the energy, {which is still well below equipartition.}

\subsubsection{Electron energization mechanisms}
\label{sec:tracingIP}

As {for the out-of-plane run} in Section~\ref{sec:tracing}, we describe electron heating and acceleration by analyzing {the behaviour} of traced particles. Fig.~\ref{fig:tracingIP} illustrates the main electron energization phases in a format similar to that in Fig.~\ref{fig:tracing}. We split the work done by the electric field into components parallel and perpendicular to the \emph{local} magnetic field (see caption of Fig.~\ref{fig:tracingIP}).

Qualitatively, with in-plane $\boldsymbol{B_0}$ the electron energization proceeds as described for $\varphi_{\rm B}=90\degree$. {Generally, particles that later gain significant energy are decoupled from the bulk flow when interacting with large-amplitude waves emitted by the ripples. An initial energy gain at time $t=t_a$ is provided by the perpendicular electric field components, $\Ex$ and the motional electric field, $\Ez$.} The magnetic moment increases. Note, that {magnetic gyration proceeds} in the $x-z$ plane, and the $\boldsymbol{E}\times\boldsymbol{B}$ drift motion due to the $\Ex$ field is {normal to the simulation plane and hence} not visible. 

\begin{figure}
\begin{center}
\includegraphics [width=0.99\linewidth] {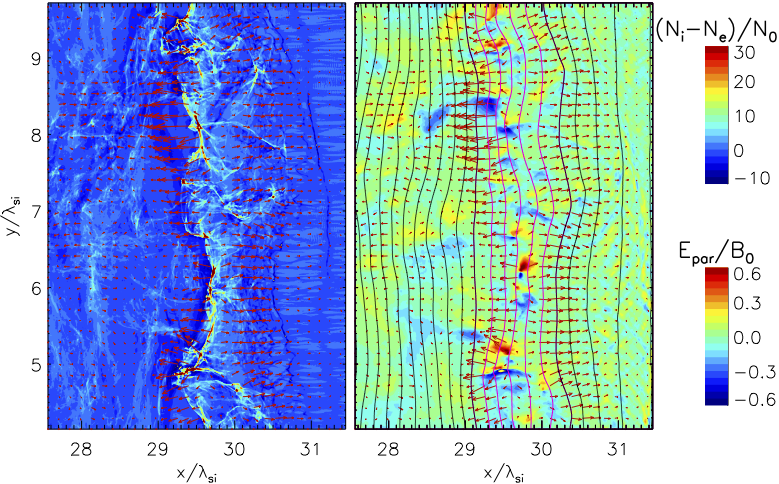}
\end{center}
\caption{\jn{Normalized charge density (left panel) and the electric-field component aligned with the magnetic field (right panel) at time $t\Wce=24.2$ in a region around the shock overshoot. Red arrows show the in-plane electric field. Magnetic field lines in the right panel are shown with black and magenta lines, the latter highlighting the region, in which $E_{\rm par}$ has a substantial amplitude.}}
\label{fig:edotb}
\end{figure}

At the shock the particles are adiabatically heated, and the magnetic moment remains roughly constant until the electrons come close to the overshoot at time $t_b$ (Fig.~\ref{fig:tracingIP}b). Then the magnetic moment starts to increase, and particles significantly gain energy as work of the $y$-component of the parallel electric field. The inset in panel (b) shows that magnetic-field-aligned electric fields exist everywhere along the magnetic overshoot. Their structure in {the larger context} of the overshoot is shown in Fig.~\ref{fig:edotb}. 

\jn{At the overshoot the density of ions is much larger than that of electrons, {on account of ion deceleration during reflection in the shock-compressed magnetic field. Charge-separation causes a strong electrostatic field associated with the cross-shock potential that reflects the ions and pulls} the electrons toward downstream. In a laminar shock the cross-shock electric field {is aligned with the shock normal, i.e., lies in $x$-direction in our setup and flips sign at the overshoot. The red arrows in Fig.~\ref{fig:edotb} in principle conform with this picture, but it is also evident that the electric field follows a} complex set of thin and highly warped filaments. The charge-separation field thus acquires a component along the magnetic field, that is largely in the $y$-direction. {The red trajectory in Fig.~\ref{fig:tracingIP}(b) gives an example of an electron that moves freely along the magnetic field and is efficiently accelerated by the aligned electric field.}}

\jn{Electrostatic field {perpendicular to the shock normal is quite common} in nonrelativistic systems. Electrons that are accelerated by the cross-shock potential can excite the two-stream instabilities \citep{Thomsen1983, Goodrich1984} {and may do so here. Fig.~\ref{fig:edotb} demonstrates modulation of the overshoot structure} by the shock ripples. The presence of modulations at scales of the ion skin depth and smaller suggests that other instabilities may also operate in the overshoot. {Disentangling the coupling between these unstable modes is difficult and beyond the} scope of this paper. We do note that the role of the $E_\parallel$ field is observable only with the in-plane magnetic-field configuration. The strength of this electric field component is much larger than that typically observed at nonrelativistic shocks {and may be specific} to mildly relativistic and magnetized shocks.}
After crossing the overshoot the average kinetic energy of the particles and the magnetic moment keep increasing, despite the magnetic-field decompression. As with out-of-plane $\boldsymbol{B_0}$, this nonadiabatic acceleration is due to gyroresonant or stochastic scattering of electrons off downstream turbulence created at shock front ripples (Fig.~\ref{fig:tracingIP}c). Note, that magnetic-field aligned electric field persists in an extended region past the overshoot.   

\section{Summary and conclusions}
This work is the second of two articles in which we use PIC simulations to investigate mildly-relativistic \jnn{superluminal} magnetized shocks in electron-ion plasma. Paper I described the electromagnetic shock structure, plasma instabilities, and waves. Here we discussed particle acceleration and heating. 

It has been suggested that wakefield acceleration at ultrarelativistic shocks may
account for the production of highly energetic particles
through nonlinear collapse of the waves \citep{lyubarsky2006,hoshino2008,iwamoto2017,iwamoto2018,iwamoto2019}. At mildly relativistic shocks, {shock ripples {at times generate strong precursor waves that lead to} nonlinear wakefield amplitudes. Then \jn{nonlinear} FRS is triggered, producing downstream-propagating wakefield that accelerates electrons through the} 
phase-slippage effect {combined with SSA} \citep{hoshino2008}. 
However, the WFA is less efficient than in the ultrarelativistic regime, {and only few particles {reach the bulk energy of upstream ions, much less exceed it. Consequently, and} in contrast to ultrarelativistic shocks, electrons and ions do not reach equipartition by the time they arrive at the shock.} 

The ion-to-electron energy transfer is comparable {for in-plane and out-of-plane magnetic field}. In both cases it is far below equipartition, with electrons on average carrying $11\%-13\%$ of the initial ion kinetic energy {or six times their \mpon{own} initial energy. Going beyond adiabatic compression at the shock, most} of the energy transfer takes place at the shock and immediately downstream. The downstream electron spectra are close to thermal distributions {with small supra-thermal components. Shock rippling is crucial for electron energization. 
Strong precursor waves emitted by the shock ripples excite wakefield that decouples particles {from the bulk flow}. They are then further accelerated in the electrostatic field at the shock. }

{With the out-of-plane magnetic field, electrons crossing the shock ramp experience adiabatic heating in the overshoot. Particles that were significantly accelerated at the rippled {shock front} now have gyro-radii comparable to scale of the downstream turbulence, and they can gain more energy through resonant wave-particle} interactions. 
Stochastic second-order Fermi-like scattering also {provides some particles with} {supra-thermal} energies. 
Similar processes shape the downstream energy spectra for the in-plane magnetic field scenario. In addition, significant energization happens at
the overshoot, where {charge-separation} {associated with the cross-shock potential and modified by the shock ripples} {reorganizes the electric field into a complex structure of warped filaments. Electric and magnetic fields partially align, leading to efficient electron} acceleration. 

We argued in Paper I that SMI-generated precursor waves should persist in realistic 3D systems and generate wakefield. We expected that the precursor waves have similar strength in 3D and in 2D, and that shock rippling operates to amplify them. {This rippling operates on small scales and is not to be confused with MHD-scale rippling that results from fluctuations in the upstream medium \citep{2016ApJ...827...44L}.}
{Similar conclusions should apply to the particle-energization efficiency. For both magnetic-field configurations {we observe similar processes that heat and accelerate electrons and thermalize the ions, although for the in-plane configuration in 2D cross-field diffusion is suppressed \citep{Jokipii1993}. The coherence of the precursor waves is likely not higher in 3D} than in lower-dimensional simulations.} 
{We conclude that the particle-acceleration efficiency is expected to be} at most the same in 3D {as} it is in 2D simulations, and we do not anticipate a stronger ion-electron coupling in three-dimensional studies of ion-electron shocks. {At ultrarelativistic \jnn{superluminal} shocks, in which the energy transfer occurs in the turbulent precursor, the level of coupling 
\jnn{increases with the shock obliquity as the strength of the precursor wave is larger for obliquities approaching a strictly perpendicular configuration
\citep{lyubarsky2006,sironi2011}.} Future studies may explore whether shock rippling can enhance the coupling at not strictly perpendicular mildly relativistic shocks.}

The absence of energy equipartition between ions and electrons has important {astrophysical} implications. {The presence of ions in blazar jets appears to be energetically required \citep{2008MNRAS.385..283C}. 
It is evident that high-energy emission from blazars requires a particle energy much in excess of that seen in our simulation. A second-stage acceleration to TeV-scale energies is required, whatever the type of radiating particle. Diffusive shock acceleration at mildly relativistic shocks is a possible process, because it can extract a large fraction of the jet energy flux and the post-shock plasma can be reasonably well confined \citep{2016RPPh...79d6901M,2017SSRv..207..319P}. In blazar jets internal shocks are magnetized and likely superluminal though, which strongly limits the efficiency of shock acceleration and suggests that other processes may be at play. Our simulations suggest that WFA is not one of them, at least for mildly relativistic systems. Weakly magnetized ultrarelativistic shocks can also not accelerate to very high energies \citep{2014MNRAS.439.2050R}. It is possible that shocks in blazar jets do not produce the high-energy particles that produce their X-ray and gamma-ray emission, and instead turbulence provides stochastic acceleration in a larger volume \citep[e.g.][]{2015MNRAS.447..530C,2016MNRAS.458.3260C}.}

{The pre-acceleration at the shock provides the initial condition for the subsequent acceleration to the TeV scale. It also shapes the electron spectra below the GeV scale, whose observational consequences can be used to make inferences about electron-ion coupling in blazar jets. If there were equipartition between ions and electrons, one would find few electrons  with Lorentz factor $\gamma_e\lesssim 10^3$. Indeed,}
internal-shock {models} for blazar jets tend to require electron-ion equipartition to reproduce blazars SEDs \citep[see, e.g.,][]{spada2001}. {It is also implicitely called for in the radiation modelling, where it appears as minimum Lorentz factor of electrons \citep{2009ApJ...692...32D,2016MNRAS.461..202A}, otherwise the models predict too much flux in X-rays and soft gamma rays.
There is explicit evidence for a low-energy cut-off at $\gamma\approx 10^4$ in the electron spectrum in the lobes of a radio galaxy \citep{2006ApJ...644L..13B}. }

{Leptonic blazar models are based on either external Compton scattering or the internal synchrotron-self-Compton process, or a combination of the two. External Compton scattering with strong electron-ion coupling, i.e. $\gamma\gtrsim 10^4$, would give X-ray spectra that are much harder than observed \citep[see, e.g.,][]{sikora2013}. Better agreement is reached, if the electrons would have cooled to low energies, resulting in a low-energy spectral tail $\propto\gamma^{-2}$, that would produce consistent X-ray spectra by synchrotron-self-Compton scattering. The energy loss time of $\gamma=100$ electrons in blazar jets is on the order of years, and so cooling tails would not be expected during flares, but rather for baseline conditions. Our simulations suggest that there is a large population of low-energy electrons on account of limited electron-ion coupling, from which particles are accelerated to very high energies. Then the low-energy part of the electron spectrum would not need cooling to form, and the observed X-ray spectra would be naturally explained.}

{Recently, a high-energy neutrino event was observed in association} with the \jn{flaring blazar} TXS 0506+056 \citep{Icecube2018}. This result requires the presence of ultrarelativistic protons in these sources and seems to exclude pure pair plasma in the jets, at least for this particular source. 
{In mixed plasma composed of ions, electrons, and positrons} an additional process may operate that accelerates electrons and positrons. Reflected ions gyrating at the shock emit left-handed elliptically polarized magnetosonic waves with frequencies that are harmonics of the ion cyclotron frequency (the ion SMI). If the spectrum of the ion-emitted waves has enough power at high harmonics at and above the cyclotron frequencies of positrons and electrons, then {positrons} traversing the shock can be {efficiently} accelerated to non-thermal energies through resonant relativistic synchrotron absorption of the magnetosonic waves \citep{hoshino1991,hoshino1992}.
Non-thermal electrons are also created, {but not as efficiently on account} of the left-handed polarization of the ion waves \citep{amato2006,stockem2012}. New {high-resolution and large-scale PIC simulations} are necessary to investigate {in 2D this scenario for mildly relativistic shocks} in ion-pair plasmas.

\section*{Acknowledgements}
\jnn{J.N. acknowledges inspiring discussions with Marek Sikora.}
This work has been supported by Narodowe Centrum Nauki through research projects DEC-2013/10/E/ST9/00662 (A.L., J.N., O.K.), UMO-2016/22/E/ST9/00061 (O.K.) and 2019/33/B/ST9/02569 (J.N.). This research was supported by PLGrid Infrastructure. Numerical experiments were conducted on the Prometheus system at ACC Cyfronet AGH. This work was supported by JSPS-PAN Bilateral Joint Research Project Grant Number 180500000671.
Part of the numerical work was conducted on resources provided by the North-German Supercomputing Alliance (HLRN) under projects bbp00003, bbp00014, and bbp00033.

\section*{Data availability}
The data underlying this article will be shared on reasonable request to the corresponding author.

\bsp	
\label{lastpage}

\begin{thebibliography}{}
\bibitem[\protect\citeauthoryear{Abdalla et al.}{2019}]{2019Natur.575..464A} Abdalla H. et al., 2019, Nature, 575, 464
\bibitem[\protect\citeauthoryear{Acciari et al.}{2019a}]{2019Natur.575..455M} Acciari V.~A. et al., 2019, Nature, 575, 455
\bibitem[\protect\citeauthoryear{Acciari et al.}{2019b}]{2019Natur.575..459M} Acciari V.~A. et al., 2019, Nature, 575, 459
\bibitem[\protect\citeauthoryear{Amato \& Arons}{2006}]{amato2006} Amato E., Arons J., 2006, ApJ, 653, 325 
\bibitem[\protect\citeauthoryear{Ansoldi et al.}{2018}]{Icecube2018} Ansoldi S. et al., 2018, ApJ, 863, L10 
\bibitem[\protect\citeauthoryear{Archambault et al.}{2016}]{2016MNRAS.461..202A} Archambault S. et al., 2016, MNRAS, 461, 202 
\bibitem[Babul \& Sironi(2020)]{2020MNRAS.499.2884B} Babul, A.-N. \& Sironi, L.\ 2020, \mnras, 499, 2884
\bibitem[\protect\citeauthoryear{Begelman \& Kirk}{1990}]{begelman1990} Begelman M.~C., Kirk J.~G., 1990, ApJ, 353, 66 
\bibitem[\protect\citeauthoryear{Blundell et al.}{2006}]{2006ApJ...644L..13B} Blundell K.~M., Fabian A.~C., Crawford C.~S., Erlund M.~C., Celotti A., 2006, ApJ, 644, L13 
\bibitem[\protect\citeauthoryear{Buneman}{1993}]{buneman1993} Buneman, O., 1993, in {Computer Space Plasma Physics: Simulation Techniques and Software}, pag.67-84 Terra Scientific Publishing Company (TERRAPUB), Tokyo
\bibitem[\protect\citeauthoryear{Burgess \& Scholer}{2007}]{burgess2007} Burgess D., Scholer M., 2007, Phys. Plasmas, 14, 012108
\bibitem[\protect\citeauthoryear{Celotti \& Ghisellini}{2008}]{2008MNRAS.385..283C} Celotti A., Ghisellini G., 2008, MNRAS, 385, 283 
\bibitem[\protect\citeauthoryear{Chen, Pohl, \& B{\"o}ttcher}{2015}]{2015MNRAS.447..530C} Chen X., Pohl M., B{\"o}ttcher M., 2015, MNRAS, 447, 530
\bibitem[\protect\citeauthoryear{Chen et al.}{2016}]{2016MNRAS.458.3260C} Chen X., Pohl M., B{\"o}ttcher M., Gao S., 2016, MNRAS, 458, 3260 
\bibitem[\protect\citeauthoryear{Dermer et al.}{2009}]{2009ApJ...692...32D} Dermer C.~D., Finke J.~D., Krug H., B{\"o}ttcher M., 2009, ApJ, 692, 32
\bibitem[\protect\citeauthoryear{Drake et al.}{1974}]{drake1974} Drake J.~F., Kaw P.~K., Lee Y.~C., Schmid G., Liu C.~S., Rosenbluth M.~N., 1974, Phys. Fluids, 17, 778
\bibitem[\protect\citeauthoryear{Gallant et al.}{1992}]{gallant1992} Gallant Y.~A., Hoshino M., Langdon A.~B., Arons J., Max C.~E., 1992, ApJ, 391, 73
\bibitem[\protect\citeauthoryear{Ghisellini, Tavecchio, \& Chiaberge}{2005}]{ghisellini2005} Ghisellini G., Tavecchio F., Chiaberge M., 2005, A\&A, 432, 401
\bibitem[\protect\citeauthoryear{Goodrich \& Scudder}{1984}]{Goodrich1984} Goodrich C.~C., Scudder J.~D., 1984, J. Geophys. Res., 89, 6654
\bibitem[\protect\citeauthoryear{Hoshino \& Arons}{1991}]{hoshino1991} Hoshino M., Arons J., 1991, Phys. Fluids B, 3, 818
\bibitem[\protect\citeauthoryear{Hoshino}{2008}]{hoshino2008} Hoshino M., 2008, ApJ, 672, 940
\bibitem[\protect\citeauthoryear{Hoshino et al.}{1992}]{hoshino1992} Hoshino M., Arons J., Gallant Y.~A., Langdon A.~B., 1992, ApJ, 390, 454
\bibitem[\protect\citeauthoryear{IceCube Collaboration et al.}{2018}]{2018Sci...361.1378I} Aartsen M.~G. et al., 2018, Science, 361, eaat1378
\bibitem[\protect\citeauthoryear{Iwamoto et al.}{2017}]{iwamoto2017} Iwamoto M., Amano T., Hoshino M., Matsumoto Y., 2017, ApJ, 840, 52
\bibitem[\protect\citeauthoryear{Iwamoto et al.}{2018}]{iwamoto2018} Iwamoto M., Amano T., Hoshino M., Matsumoto Y., 2018, ApJ, 858, 93
\bibitem[\protect\citeauthoryear{Iwamoto et al.}{2019}]{iwamoto2019} Iwamoto M., Amano T., Hoshino M., Matsumoto Y., Niemiec J., Ligorini A., Kobzar O., Pohl, M., 2019, ApJ, 883, L35
\bibitem[\protect\citeauthoryear{Jokipii, Kota, \& Giacalone}{1993}]{Jokipii1993} Jokipii J.~R., Kota J., Giacalone J., 1993, Geophys. Res. Lett., 20, 1759
\bibitem[\protect\citeauthoryear{Kaw, Schmidt, \& Wilcox}{1973}]{kaw1973} Kaw P., Schmidt G., Wilcox T., 1973, Phys. Fluids, 16, 1522
\bibitem[\protect\citeauthoryear{Kruer}{1988}]{kruer1988} Kruer W.~L., 1988, The physics of laser plasma interactions, Reading, MA, Addison-Wesley Publishing Co.
\bibitem[\protect\citeauthoryear{Kuramitsu et al.}{2008}]{kuramitsu2008} Kuramitsu Y., Sakawa Y., Kato T., Takabe H., Hoshino M., 2008, ApJ, 682, L113
\bibitem[\protect\citeauthoryear{Langdon, Arons, \& Max}{1988}]{langdon1988} Langdon A.~B., Arons J., Max C.~E., 1988, Phys. Rev. Lett., 61, 779
\bibitem[\protect\citeauthoryear{Lemoine, Ramos, \& Gremillet}{2016}]{2016ApJ...827...44L} Lemoine M., Ramos O., Gremillet L., 2016, ApJ, 827, 44
\bibitem[\protect\citeauthoryear{Ligorini et al.}{2021}]{2020MNRAS.tmp.3691L} Ligorini A., Niemiec J., Kobzar O., Iwamoto M., Bohdan A., Pohl M., Matsumoto Y., et al., 2021, MNRAS, 501, 4837
\bibitem[\protect\citeauthoryear{Lyubarsky}{2006}]{lyubarsky2006} Lyubarsky Y., 2006, ApJ, 652, 1297
\bibitem[\protect\citeauthoryear{Marcowith et al.}{2016}]{2016RPPh...79d6901M} Marcowith A., Bret A., Bykov A., Dieckman M.~E., O'C Drury L., Lemb{\`e}ge B., Lemoine M., et al., 2016, Rep. Prog. Phys., 79, 046901
\bibitem[\protect\citeauthoryear{Martins et al.}{2009}]{2009ApJ...695L.189M} Martins S.~F., Fonseca R.~A., Silva L.~O., Mori W.~B., 2009, ApJ, 695, L189
\bibitem[\protect\citeauthoryear{Niemiec, Ostrowski, \& Pohl}{2006}]{2006ApJ...650.1020N} Niemiec J., Ostrowski M., Pohl M., 2006, ApJ, 650, 1020
\bibitem[\protect\citeauthoryear{Niemiec et al.}{2008}]{niemiec2008} Niemiec J., Pohl M., Stroman T., Nishikawa K.-I., 2008, ApJ, 684, 1174
\bibitem[\protect\citeauthoryear{Pelletier et al.}{2017}]{2017SSRv..207..319P} Pelletier G., Bykov A., Ellison D., Lemoine M., 2017, Space Sci. Rev., 207, 319
\bibitem[\protect\citeauthoryear{Plotnikov, Grassi, \& Grech}{2018}]{plotnikov2018} Plotnikov I., Grassi A., Grech M., 2018, MNRAS, 477, 5238
\bibitem[\protect\citeauthoryear{Plotnikov \& Sironi}{2019}]{plotnikov2019} Plotnikov I., Sironi L., 2019, MNRAS, 485, 3816
\bibitem[\protect\citeauthoryear{Pohl, Hoshino, \& Niemiec}{2020}]{2020PrPNP.11103751P} Pohl M., Hoshino M., Niemiec J., 2020, Prog. Part. Nucl. Phys., 111, 103751
\bibitem[\protect\citeauthoryear{Reville \& Bell}{2014}]{2014MNRAS.439.2050R} Reville B., Bell A.~R., 2014, MNRAS, 439, 2050
\bibitem[\protect\citeauthoryear{Sikora et al.}{2013}]{sikora2013} Sikora M., Janiak M., Nalewajko K., Madejski G.~M., Moderski R., 2013, ApJ, 779, 68
\bibitem[\protect\citeauthoryear{Sironi \& Spitkovsky}{2009}]{sironi2009} Sironi L., Spitkovsky A., 2009, ApJ, 698, 1523
\bibitem[\protect\citeauthoryear{Sironi \& Spitkovsky}{2011}]{sironi2011} Sironi L., Spitkovsky A., 2011, ApJ, 726, 75
\bibitem[\protect\citeauthoryear{Spada et al.}{2001}]{spada2001} Spada M., Ghisellini G., Lazzati D., Celotti A., 2001, MNRAS, 325, 1559
\bibitem[\protect\citeauthoryear{Stockem et al.}{2012}]{stockem2012} Stockem A., Fi{\'u}za F., Fonseca R.~A., Silva L.~O., 2012, ApJ, 755, 68
\bibitem[\protect\citeauthoryear{Tajima \& Dawson}{1979}]{1979PhRvL..43..267T} Tajima T., Dawson J.~M., 1979, Phys. Rev. Lett., 43, 267
\bibitem[\protect\citeauthoryear{Thomsen et al.}{1983}]{Thomsen1983} Thomsen M.~F., Barr H.~C., Gary S.~P., Feldman W.~C., Cole T.~E., 1983, J. Geophys. Res., 88, 3035
\end{thebibliography}
\end{document}